\documentstyle[twoside,fleqn,espcrc2,epsf]{article}

\title{
\vspace{-6.00cm}
\begin{flushright}
{\normalsize FSU-SCRI-97-97}\\
\vspace{-0.3cm}
{\normalsize UW/PT 97-21}\\
\vspace{-0.1cm}
hep-lat/9709014\\
\end{flushright}
\vspace{2.5cm}
The Overlap Formalism and Topological Susceptibility on the Lattice
\thanks{This work is based on a talk given by Robert Singleton at
Lattice '97, held in Edinburgh, Scotland.}
}

\author{
Rajamani Narayanan\address{SCRI, 
The Florida State University, Tallahassee, FL 32306, USA\\
rnaray@scri.fsu.edu}
\thanks{This research was supported by DOE contracts 
DE-FG05-85ER250000 and DE-FG05-96ER40979.}
  and 
Robert L. Singleton Jr.\address{Department of Physics,
The University of Washington, Seattle, WA 98195, USA\\
bobs@terrapin.phys.washington.edu}
\thanks{This work was supported by DOE
contract DE-FG03-96ER40956.}
}
       
\begin{document}

\begin{abstract}
Recent lattice measurements of the topological susceptibility
of $SU(2)$ gauge theory using improved cooling and inverse-blocking 
are in disagreement. We use the overlap method, which probes 
the fermionic sector of the theory directly, to help resolve 
this discrepancy.
\end{abstract}

\maketitle

Two recent methods\cite{defps,dhzk} for measuring the topological 
susceptibility, $\chi_t = {\langle Q^2 \rangle\over V}$, in pure 
$SU(2)$ gauge theory using the lattice regularization are considered 
to be in disagreement. In one method\cite{defps}, the gauge fields 
are smoothed with an improved cooling technique while the topological 
charge $Q$ is calculated using a lattice discretization. In the other 
method\cite{dhzk}, inverse-blocking is employed to smooth the lattice 
configurations, after which the topological charge is calculated using 
an ``algebraic'' definition. In both methods, scaling is assumed and 
a continuum limit for $\chi_t$  obtained, with the value of 
Ref.~\cite{defps} being about 20\% smaller than that of Ref.~\cite{dhzk}. 
It should be noted that the standard Wilson plaquette action is used 
in the first method to generate the gauge field configurations on the 
lattice, while in the second method the gauge field ensemble is 
generated with a certain fixed-point action. Therefore, in addition 
to scaling, it is also necessary to assume universality to compare 
the two numbers above.

The index of the Euclidean chiral Dirac operator on the lattice
may be extracted by the overlap method\cite{nn} in a clean manner,
that is to say, in a way that does not require the smoothing of 
background gauge fields. As the Atiyah-Singer index 
theorem\cite{asit} equates the index to the topological charge, 
the overlap method should thus be useful in probing for topological 
structure. One should keep in mind that the index theorem is a 
statement about continuum gauge fields and its validity for
configurations on the lattice should first be tested.  Ref.~\cite{nv} 
measures the index distribution of the discretized chiral Dirac 
operator in pure $SU(2)$ gauge theory using the standard Wilson action 
at a coupling of $\beta=2.4$ on a $12^4$ lattice, which enables a direct 
comparison with the distribution of topological charge in Ref.~\cite{defps}. 
The index distribution matches extremely well with the topological charge 
distribution in accordance with the index theorem. This indicates that we 
may in fact apply the index theorem to an ensemble of lattice gauge field 
configurations, even when these configurations themselves are not smooth. 
Furthermore, the agreement between the index distribution from the overlap 
method and the distribution of topological charge using improved cooling 
suggests that overlap techniques might help resolve the discrepancy 
stated in the first paragraph. 

\begin{figure}[t]
\vskip0.3cm
\epsfxsize=3.0in
\centerline{\epsffile{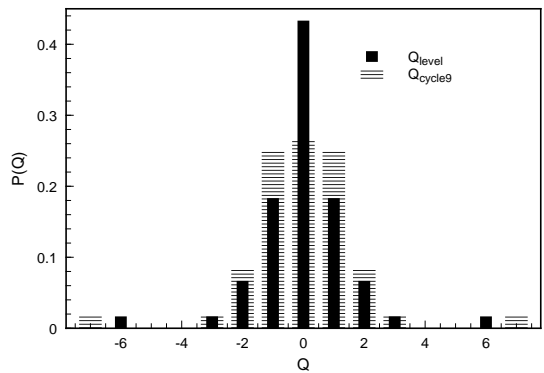}}
\caption{The index distribution measured by the overlap 
method, $Q_{\rm level}$, compared against the topological 
charge distribution obtained by inverse-blocking, 
$Q_{\rm cycle9}$, on the same ensemble of configurations. 
For the latter, 9 cycles of inverse-blocking were employed.}
\label{fig:hist}
\end{figure}

To this end, Anna Hasenfratz provided us with 30 configurations 
generated on a $12^4$ lattice using the fixed point action of 
Ref.~\cite{dhzk}. In Fig.~\ref{fig:hist}, we compare the 
corresponding topological charge distribution as measured by 
inverse-blocking with the index distribution of the chiral 
Dirac operator obtained by overlap techniques. Clearly the 
broader distribution of topological charge arising from 
inverse-blocking is inconsistent with the index distribution 
measured by the overlap. Furthermore, if one had estimated 
the topological susceptibility using the latter, the result 
would be smaller than that obtained by the inverse-blocking 
method of Ref.~\cite{dhzk}, and possibly closer to the improved 
cooling measurement of Ref.~\cite{defps}. To gain additional 
insight into the difference between the two distributions in 
Fig.~\ref{fig:hist}, we must examine the inverse-blocking 
procedure in more detail.

Inverse-blocking is used by Ref.~\cite{dhzk} to smooth the 
gauge fields so that topological objects can be resolved from 
the rough thermal backgrounds. An original configuration on the
$12^4$ lattice is mapped to a configuration on a $24^4$ lattice 
by an inverse-blocking transformation, which results in a smoother 
configuration on the finer lattice. In practice, however, one 
inverse-blocking step does not sufficiently smooth the configuration, 
and one must therefore go to even finer lattices. But this is not 
feasible with present day computers, and the following alternative 
is employed instead. After the rough configuration on the original 
lattice is inverse-blocked to the finer lattice, it is subsequently 
blocked back to the coarser lattice, after which it is expected to 
remain just as smooth as it was on the finer lattice. If desired,
this procedure can be repeated again on the resulting configuration.
Successive cycles of inverse-blocking followed by blocking will 
smooth the configuration a little more each time, until eventually 
topology can be resolved. To obtain the topological charge distribution 
of Fig.~\ref{fig:hist}, for example, each configuration of the ensemble 
was cycled nine times, after which the topological charge was measured 
using an ``algebraic'' topological charge operator.

The procedures of Refs.~\cite{defps} and \cite{dhzk}, in which gauge 
field smoothing is an intricate part of the measurement process, 
should be contrasted with overlap measurements, which can be performed 
on the original unsmoothed configurations. The index of the chiral Dirac 
operator is simply given by the number of level crossings of the 
Hermitian Hamiltonian $H(\mu)=\gamma_5 W(\kappa)$, where $W(\kappa)$ 
is the usual Wilson Dirac operator, and the mass parameter 
$\mu=4-{1\over 2\kappa}$ can range over all real values. For example, 
the index distribution of Fig.~\ref{fig:hist} was determined by 
counting the number of net level crossings associated with each 
configuration of the original unsmoothed ensemble. The connection 
between level crossing and the index is intimately tied to the 
definition of the chiral determinant, and the details can be found 
in \cite{nn} and \cite{nv}. 

\begin{figure}[t]
\epsfxsize=3.0in
\centerline{\epsffile{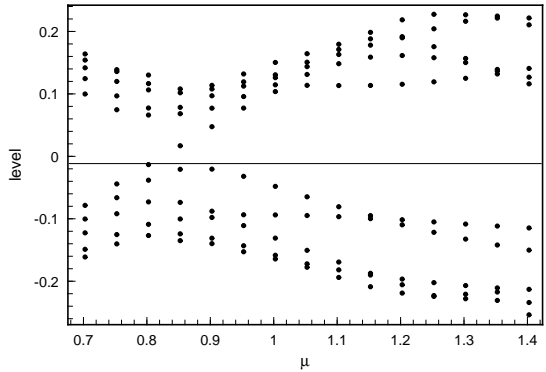}}
\caption{Level crossing for a specific uncylced configuration.}
\label{fig:lvcf6}
\end{figure}

\begin{figure}[t]
\epsfxsize=3.0in
\centerline{\epsffile{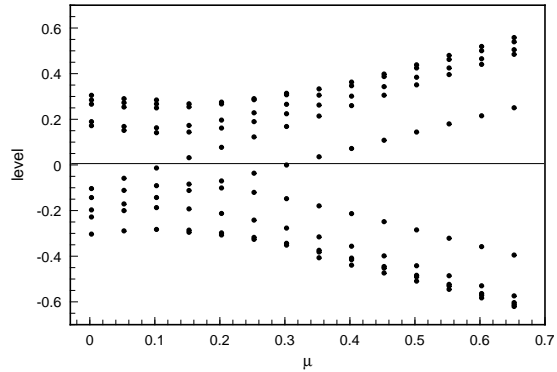}}
\caption{Level crossing for the configuration of Fig.~2
after 12 cycles.}
\label{fig:lvcf6cy}
\end{figure}

To understand the origin of the discrepancy between the two 
distributions in Fig.~\ref{fig:hist}, it is useful to focus
on individual configurations. Fig.~\ref{fig:lvcf6} illustrates
the spectral flow associated with a specific gauge field of
the original unsmoothed ensemble, and we see that the overlap 
method gives an index equal to one for the chiral Dirac operator. 
This should be contrasted with the inverse-blocking method, 
which measures the topology of this configuration to be three 
after 9 cycles and two after 12 cycles. The overlap method can 
also be used to measure the index on cycled configurations 
themselves, which provides additional insight into the effects 
of cycling. The spectral flow of the 12-cycled configuration 
is plotted in Fig.~\ref{fig:lvcf6cy}, which shows a net crossing 
of two, in agreement with the inverse-blocking measurement on 
the same configuration;\footnote{It is likely that the overlap 
measurement on the 9-cycled configuration would also agree with 
the inverse-blocking result, but the 9-cycled configuration was 
not readily available and we were unable to perform the measurement.} 
however, the aim is to determine the topology of the original 
configuration. The success of the inverse-blocking scheme depends 
on its ability to preserve the topology as the configuration is 
cycled, and this example shows that this is not always the case. 


The overlap method can also be used to obtain the size of topological
objects. In fact, one can show that the levels of larger instantons 
cross at smaller values of $\mu$, and hence measuring the crossing-point 
can provide a rough indication of the instanton size. Is it also possible 
to get a much more precise determination of the instanton size directly 
from the corresponding eigenvectors near the crossing point. This work 
is in progress and should reveal even more detailed comparisons of 
Figs.~\ref{fig:lvcf6} and \ref{fig:lvcf6cy}.

In closing, the overlap method indicates that the cycling procedure 
of Ref.~\cite{dhzk} has not only smoothed the configuration studied
above,\footnote{Note that the level crossing in Fig.~\ref{fig:lvcf6cy}
occurs at a much smaller value of $\mu$ than in the corresponding 
Fig.~\ref{fig:lvcf6}, which indeed shows that cycling has smoothed 
the configuration considerably.} but it has also changed the topology 
by generating an instanton. If cycling can increase the index, as the 
example suggests, it is conceivable that this is the cause for the 
broader distribution in Fig.~\ref{fig:hist} and an explanation of 
the disagreement stated at the beginning of this paper. 

\section*{Acknowledgments}
We would like to thank Robert Edwards, Anna Hasenfratz, Urs Heller, 
and T\'{a}mas Kov\'{a}cs for several discussions. We would also like 
to thank Anna Hasenfratz for providing us with a fixed-point ensemble 
and the associated topological charges used for Fig.~\ref{fig:hist}, 
and also for providing us with a 12-cycled configuration. 
R.N. would like to acknowledge the
Aspen Center for Physics where much of this work was carried 
out. Computations resulting in the spectral flow were performed 
on the CM-2 at SCRI.

\end{document}